\documentclass[12pt]{elsarticle}

\usepackage{natbib}
\usepackage{amssymb}
\usepackage{amsmath}
\usepackage{graphicx}
\usepackage{bm}
\usepackage{epsfig}
\usepackage{textcomp}
\biboptions{sort&compress}
\begin{document}
\begin{frontmatter}
\title{Machine learning approach for the inverse problem and unfolding procedures}
\author{N.D. Gagunashvili\corref{cor1}}
\ead{nikolai@unak.is}
\cortext[cor1]{Tel.: +354-4608505; fax: +354-4608998}
\address{University of Akureyri, Borgir, v/Nordursl\'od, IS-600 Akureyri, Iceland}

\begin{abstract}
A procedure for unfolding the true distribution from experimental data is presented. Machine learning methods are applied for simultaneous identification of an apparatus function and solving of an inverse problem. A priori information about the true distribution from theory or previous experiments is used for Monte-Carlo simulation of the training sample. The training sample can be used to calculate a transformation from the true distribution to the measured one. This transformation provides a robust solution for an unfolding problem with minimal biases and statistical errors for the set of distributions used to create the training sample. The dimensionality of the solved problem can be arbitrary. A numerical example is presented to illustrate and validate the procedure.
\end{abstract}
\begin{keyword}
unfolding \sep system identification \sep D-optimization \sep apparatus function \sep deconvolution \sep robustness \sep boosting
\PACS 02.30.Zz \sep 07.05.Kf \sep 07.05.Fb
\end{keyword}
\end{frontmatter}

\section{Introduction}
An experimentally measured distribution differs from the true physical distribution because of the limited efficiency of event registration and the finite resolution of a particular set-up. To identify a physical distribution, an unfolding procedure is typically applied \cite{zhigunov2, blobel, correcting, gagunashvili,schmelling,hocker,zech,agost, blobel2, gagunashvili_phystat,albert}. Unfolding is an underspecified problem. Any approach to solving the problem requires a priori information about the solution. Methods for unfolding differ, directly or indirectly, in the use of this a priori information.

Unfolding when the apparatus function or transformation model for a true distribution from the measured one is unknown has been considered previously \cite {gagunashvili, gagunashvili_phystat}. In this paper these ideas are further developed and the problem of simultaneously identifying a transformation model and inverse problem is solved. To obtain a robust solution for an unfolding problem, information about the shape of the distribution to be measured is used to create a training sample in Monte-Carlo simulations of an experiment.
An approximation of the apparatus function is calculated for the set of distributions for the training sample. Use of this type approximation can minimize the statistical errors and biases of the unfolded distribution for  distributions used to create the training sample.  There is no restriction on the size and shape of bins, linearization of the problem is simple (if the set-up has non-linear distortions), and multidimensional data can be unfolded.
A machine learning approach provides a method for validating the unfolding procedure and for improving the results.

The remainder of the paper is organized as follows. In Section 2 the main equation for solving an unfolding problem is proposed. A formal method for solving the unfolding problem and estimating the statistical errors for the unfolded distribution is discussed. Section 3 presents the algorithm for calculating the transformation matrix. In Section 4 the overall unfolding procedure is described. This consists of bin choice, system identification, solution of the basic equation and validation  of the unfolding procedure. Section 5 presents a numerical example. For comparison, an example reported elsewhere is used \cite{blobel, schmelling, hocker}. To investigate biases  in the unfolding distribution, a numerical experiment with 1000 runs is performed. The results show that biases for the unfolded distribution is small. To demonstrate the robustness of the unfolding method for  distributions used to create the training sample, the same investigation is performed for eight distributions randomly chosen from training sample. The results reveal that there are small biases and low statistical errors for all the unfolding distributions, which confirms that the procedure is robust. Statistical errors are as small as possible in all cases because of application of the least mean square method and the method for system identification.

\section{Main equation}
In this work we use a linear model to transform a true distribution to the measured one:
\begin{equation}
\bm{f}=\mathsf{P}\bm{\phi}+\bm{\epsilon}\, , \label{basic}
\end{equation}
where $\bm{f}$ is an $m$-component column vector of an experimentally measured histogram, ${\mathsf{P}}$ is an $m \times n$ matrix, with $m \ge n$, $\bm{\phi}$ is an $n$-component vector of some true histogram and $\bm{\epsilon}$ is an $m$-component vector of random residuals with expectation value $\mathrm{E}\,\bm{\epsilon}=\bm{0}$ and a diagonal variance matrix $\mathsf{\Sigma}=\mathrm{Var}\,\bm{\epsilon} =\mathrm{diag} (\sigma_1^2, \sigma_2^2, \cdots ,\sigma_m^2)$, where $\sigma_i$ is the statistical error of the measured distribution for the $i$th bin.
The linear model (\ref{basic}) is reasonable for the majority of set-ups in particle and nuclear physics. It is only an approximate model for set-ups with a non-linear transformation from a true to a measured distribution.

A least squares method \cite{seber} can give an estimate of the true distribution ${\bm{\phi}}$,
\begin{equation}
\hat{\bm{\phi}}=(\mathsf{P}'\mathsf{\Sigma}^{-1}\mathsf{P})^{-1}\mathsf{P}'\mathsf{\Sigma}^{-1}\bm{f} ,
\label{unfolded}
\end{equation}
where $\hat{\bm{\phi}}$, the estimate, is the {\em unfolded distribution} and the variance matrix of the unfolded distribution $\mathsf{\Delta}$ is given by
\begin{equation}
\mathsf{\Delta}=\mathrm{Var}\,\hat{\bm{\phi}}=(\mathsf{P}'\mathsf{\Sigma}^{-1}\mathsf{P})^{-1} . \label{error}
\end{equation}
The diagonal element $\delta_{ii}^2$ of the matrix is the variance of component $\hat {\phi_i}$ of the unfolded vector and $\delta_{ii}$ is the statistical error.
\section{Identification of the transformation model}
To realize the scheme described in Section 2, the matrix $\mathsf{P}$ must be defined. This problem can be solved using system identification methods \cite{graupe,ljung}. System identification can be defined as a process for determining a model of a dynamic system using observed input--output data. In our case, this is the model for transforming a true physical distribution into the experimentally measured distribution, represented by the matrix $\mathsf{P}$. Monte-Carlo simulation of a set-up can be used to obtain input--output data. Control input signals are used for system identification. The most popular choice is to use impulse control signals \cite{graupe,ljung}.

An impulse input control signal is a generated (input) distribution in which the histogram with $n$ bins has only one bin with non-zero content. For model (\ref{basic}), there are $n$ different impulse inputs that can be presented as the diagonal matrix $\mathsf{\Phi^{c}}=\mathrm{diag}(\phi_{11}^c, \phi_{22}^c, \dots ,\phi_{nn}^c)$, where each row contains the content from a generated histogram. Denote the corresponding values of the $i$th component of the reconstructed (output) vector as $\bm{f_i^c}=(f_{i1}^c f_{i2}^c \cdots f_{im}^c)'$. Each element of the $i$th row of the matrix
\[ \mathsf{P} = \left(\begin{array}{llcl}
 p_{11} & p_{12} & \cdots & p_{1n} \\
 \multicolumn{4}{c}\dotfill\\
 p_{i1} & p_{i2} & \cdots & p_{in} \\
 \multicolumn{4}{c}\dotfill\\
 p_{m1} & p_{m2} & \cdots & p_{mn} \\
 \end{array}\right)\, \]
can be found from the equation
\begin{equation}
 \bm{f_i^c}=\mathsf{\Phi^{c}}\bm{p_i}\, ,
\end{equation}
 where $\bm{p_i}=( p_{i1} \, p_{i2} \cdots p_{in})'\, $, and $p_{ij}=f_{ij}^c/\phi_{jj}$.
Equation (\ref{unfolded}), with the matrix $\mathsf{P}$ calculated in this way, gives a highly fluctuating unfolded function with large statistical errors. In addition, it is possible that the matrix $\mathsf{P}'\mathsf{\Sigma}^{-1}\mathsf{P}$ is singular, in which case a solution does not exist. The effect of this type of instability is well known. There are many methods for solving this type of system, all of which use a priori information to obtain a stable solution to Eq. (\ref{basic}).

For system identification, instead of using impulse control distributions, we use a training sample of distributions based on a priori physically motivated information that may be known from theory or from some other experimental data.

Assume that we have a training sample with $q$ generated (input) distributions and presented as a $q \times n$ matrix
\[ \mathsf{\Phi^{c}} = \left(\begin{array}{llcl}
 \phi_{11}^c & \phi_{12}^c & \cdots & \phi_{1n}^c \\
 \phi_{21}^c & \phi_{22}^c & \cdots & \phi_{2n}^c \\
 \multicolumn{4}{c}\dotfill\\
 \phi_{q1}^c & \phi_{q2}^c & \cdots & \phi_{qn}^c \\
 \end{array}\right) \, , \] \label{matrix}
where each row represents a generated histogram content. For each $i$th row of the matrix $\mathsf{P}$, we can write the following equation \cite{graupe}:
\begin{equation}
\bm{f_i^c}=\mathsf{\Phi^c}\bm {p_i}+\bm{\xi_i}\, , \label{ident}
\end{equation}
where $\bm{p_i}=( p_{i1} \, p_{i2} \cdots p_{in})'$, $\bm{f_i^c}$ is a $q$-component vector of the content for the reconstructed (output) $i$th bin for different generated distributions, and $\bm{\xi_i}$ is a $q$-component vector of random residuals with expectation value $\mathrm{E} \bm{\xi_i}=\bm{0}$ and a diagonal variance matrix $\mathsf{\Gamma_i}=\mathrm{Var} \,\bm{\xi_i} =\mathrm{diag}(\gamma_{i1}^2, \cdots , \gamma_{iq}^2)$, where $\gamma_{ij}$ is the statistical error for the reconstructed distribution for the $i$th bin and the $j$th generated distribution.
Formally a least squares method gives an estimate for $\bm{p_i}, i=1, \dots ,m$:
\begin{equation}
\hat{\bm{p_i}}=(\mathsf{\Phi^c}'\mathsf{\Gamma_i}^{-1}\mathsf{\Phi^c})^{-1}\mathsf{\Phi^c}'\mathsf{\Gamma_i}^{-1}\bm{f_i^c} \, . \label{pmatr}
\end{equation}
The whole matrix $P$ is found by producing calculations defined by formula (\ref{pmatr}) for all rows.

Similarity of shapes of distributions of the training sample leads to high correlations between columns of matrix $\mathsf{\Phi^{c}}$. This means that transformation of generated distribution to the $i$th bin of the reconstructed distribution can be parameterized using the subset of elements of row $\bm{p_i}$.  Elements of a row that do not belong to the subset are set to 0.

The training sample contained copies of the same distribution is example of the singular case of the similarity. The transformation can be reduced to only one non-zero element of   vector $\hat{\bm{p_i}}$ for this example.

Another example is the training sample that contains  any possible distributions. The number of non-zero elements cannot  be reduced and  matrix $\mathsf{P}$  coincides with matrix calculated using impulse control signals.

A forward stepwise regression algorithm can be used \cite{seber} to find non-zero elements of a row $\bm{p_i}$.
Stepwise algorithm combines FS and BE steps. Steps  are followed by each other and repeated until the process is terminated. Steps are defined as:\\
\textbf{Step FS.} Suppose there is $l$ elements of row $i$ included into the model of transformation. Subvector of elements $\bm {p_{i}(l)}$ is calculated according to formula (\ref{pmatr})  with submatrixes $ \mathsf{\Phi^c\mathbf{}}(l)$ and $\mathsf{\Gamma_i}(l)$ that correspond to this subvector.  A new element is  added if:
\begin{equation}
\frac{X_l^2-X_{l+1}^2}{X_{l+1}^2}(n-l-1)>F_{in} \label{eneq}
\end{equation}
where
\begin{equation}
X_{l}^2= [\bm{f_i^c}-\mathsf{\Phi^c\mathbf{}}(l)\bm {p_{i}(l)}]'\,\, \mathsf{\Gamma_i}^{-1}(l)\,\,[\bm{f_i^c}-\mathsf{\Phi^c\mathbf{}}(l)\bm {p_{i}(l)}]
\end{equation}
and
$F_{in}$ is constant (threshold). \\
\textbf{Step BE.} Let there be $l$ elements of row $i$ included into model of transformation then an element is excluded from model of transformation  if:
\begin{equation}
\frac{X_{l-1}^2-X_{l}^2}{X_{l}^2}(n-l)< F_{out} \label{eneqout}
\end{equation}
where $F_{out}$ is another constant.

The algorithm is terminated when there cannot be found any elements that satisfy
inequality (\ref{eneq}) or inequality (\ref{eneqout}). Good results give thresholds $F_{in}=F_{out}=3.29$ that have some theoretical background, see \cite{draper}.
Position $k$ of the first element $p_{ik}$  for our case is defined by the maximum value of the correlation between vector $\bm{f_i^c}$ and columns of matrix $\mathsf{\Phi^{c}}$:
\begin{equation}
\mathrm{Cor}(\bm{f_i^c}, \bm{\phi_{ k}^c})= \max [\mathrm{Cor}(\bm {f_i^c}, \bm{\phi_{1}^c)}, \mathrm{Cor}(\bm{f_i^c},\bm{\phi_{ 2}^c}),\ldots , \mathrm{Cor}(\bm{f_i^c},\bm{\phi_{ n}^c})] ,
\end{equation}
where $\bm{\phi_{j}^c}=(\phi_{1j}^c \phi_{2j}^c \ldots \phi_{qj}^c)'$.

The whole matrix $P$ is found by stepwise algorithm calculations for all rows.

It is possible that for each row exist more than one subset of non-zero matrix elements that describe the transformation in a sufficiently good manner.
This case  can be, for example, when all distributions of training sample are rather close to each other.
 Thus, for each $i$th reconstructed bin we will have a set of $N_i$ candidate rows, and for all reconstructed bins a set of $N_1 \times N_2 \times \cdots \times N_m$ candidate matrices $\mathsf{P}$. We need to choose a matrix $\mathsf{P}$ that is good or optimal in some sense. The most convenient criterion in our case is $\mathrm{D}$-optimality \cite{fedorov}, which is related to minimization of
\begin{equation}
\mathrm{det}(\mathsf{P}'\mathsf{\Sigma}^{-1}\mathsf{P})^{-1}=\mathrm{det}(\mathrm{Var}\,(\bm{\hat{\phi}}))\, .\label{det}
\end{equation}
There are many algorithms and programs for minimization of (\ref{det}). The matrix $\mathsf{P}$ that minimizes function (\ref{det}) gives a stable solution to unfolding problem (\ref{unfolded}) with a minimum volume for the confidence ellipsoid.

There are three possibilities to further improve the quality of the solution:
\begin{enumerate}

\item Introduce selection criteria for models of distributions used to create the training sample. The previously described goodness-of-fit test can be used for this purpose \cite {goodness}.

\item Each training distribution has a reconstructed distribution that can be compared with the experimentally measured distribution using a $\chi^2$ test \cite{gagunashviliph}. Improvement is achieved by selecting distributions for the training sample that satisfy $X^2<a$, where $X^2$ is the test statistic for comparison of the reconstructed and experimentally measured distributions \cite{gagunashvili_phystat}. The threshold $a$ defines how close a reconstructed distribution is to the experimental distribution. Note that any threshold $a$ corresponds to a particular significance level for the test. It is reasonable that a decrease in parameter $a$ represents a decrease in bias and statistical error for the solution.

\item A leave-one-out validation procedure \cite {tan,weka} for $q$ runs can be performed. During a run the unfolding procedure is applied for each of $q$ a reconstructed distributions. Each unfolded distribution is then compared with the corresponding generated distribution using a $\chi^2$ test \cite{seber}. A boosting procedure \cite {tan,weka} can be used for distributions of the training sample with a low $p$-value. This involves adding to the training sample the same distribution with a statistically independent realization of the corresponding reconstructed histogram.

\end{enumerate}
\section{Unfolding procedure}
This section provides a description of the complete unfolding procedure. The procedure can be divided into four parts: initialization, system identification, solution of the basic equation, and validation.\\

\noindent
{\bf Initialization}

\noindent
\begin{itemize}
\item {\em Define the binning for the experimental (reconstructed) data.}
The strategy for selecting the bin size involves starting with a large bin size and then increasing the number of bins incrementally until the error for the unfolded distribution stops decreasing.

\item {\em Define the binning for the unfolded (generated) distribution.} The bin size should be chosen by picking a reasonably large size first and then decreasing the size in further steps until the correlation between adjusted bins becomes too large. The number of bins for an unfolded distribution, $n$, must be less than the number of bins for the experimentally measured distribution, $m$, because the least squares method is used to solve the main equation.
 \end{itemize}
\newpage
\noindent
{\bf System identification}
\begin {itemize}
\item {\em Choose a training sample of generated distributions.}
Generated distributions for the training sample must be chosen as described in the previous section. A second iteration can be made to find a better set of distributions. The number of generated distributions must be greater than the expected number of non-zero elements in any row of matrix $P$ (for reasons related to use of the least squares method).

\item {\em Calculate the matrix} $\mathsf{P}$. The matrix is calculated according to the algorithm described in the previous section.

\item {\em Calculate the D-optimal matrix} $\mathsf{P}$. Optimization can be performed using Fedorov's reliable EA algorithm \cite{fedorov} with the initial matrix $P$ calculated in the previous step.
\end {itemize}
\noindent
{\bf Solution of the basic equation}
\begin {itemize}
\item {\em Calculate the unfolded distribution Eq. (\ref{unfolded}) with the variance matrix Eq. (\ref{error})}. The correlation matrix calculated from the variance matrix can give hints for improved binning of the unfolding distribution. For example, if the correlation between two adjacent bins is high, they should be combined.
\end {itemize}
\noindent
{\bf Validation of the unfolding procedure}
\begin{itemize}
\item {\em Fit the unfolded distribution, and then use this fit to generate a reconstructed distribution (including the effects of resolution and acceptance) to compare with the real data.}

\item {\em Leave-one-out procedure \cite {tan,weka} for $q$ runs. During a run, the unfolded procedure is applied for each of $q$ reconstructed distributions.} The unfolded distributions are then compared with the corresponding generated distributions \cite {seber}.

\end{itemize}

This procedure yields an unfolded distribution with minimal statistical errors and minimal bias for the true distributions closed to distributions of the training sample. This follows from the properties of the least mean square method and the method used to calculate the transformation matrix $P$.

\section{A numerical example}
The method described above is now illustrated with an example proposed by Blobel \cite {blobel} and used for illustration elsewhere \cite{schmelling,hocker}.
We take a true distribution
\begin{equation}
\phi(x) \propto \sum^3_{i=1} A_i\frac{C_i^2}{(x-B_i)^2+C_i^2} \label{testform}
\end{equation}
with the same parameters as in a previous study \cite{blobel} (Table 1, first row), where $x$ is defined on the interval $[0, 2]$.\\
\noindent
\vspace *{-.5 cm}
\begin{table}[h]
\centering
\caption{Values of the parameters and intervals used for training sample simulations}
\vspace *{0.5cm}
\footnotesize
\hspace *{-0.0cm}\begin{tabular}{c|c|c|c|c|c|c|c|c}
$A_1$ & $A_2$ & $A_3$ & $B_1$ & $B_2$ & $B_3$ & $C_1$ & $C_2$ & $C_3$ \\
\hline
$1$ & $10$ & $5$ & $0.4$ & $0.8$ & $1.5$ & $2$ & $0.2$ & $0.2$\\
$[0.5,3]$ & $[6,14]$& $[1,9]$& $[0.2,0.6]$ & $[0.6,1.3]$ & $[1.3,2]$ & $[0.5,3.5]$ &$[0.1,0.4]$& $[0.1,0.5]$ \\
\end{tabular}
\end{table}

\noindent
An experimentally measured distribution is defined as
\begin{equation}
f(x)=\int_0^{2} \phi(x')A(x')R(x,x')dx' \,
\end{equation}
where the acceptance function $A(x)$ is
\begin{equation}
A(x)=1-\frac{(x-1)^2}{2}
\end{equation}
 and
\begin{equation}
R(x,x')=\frac{1}{\sqrt{2\mathrm{\pi}}\sigma}\mathrm{exp}(-\frac{(x-x'+0.05x'^2)^2}{2\sigma^2})
\end{equation}
is the detector resolution function with $\sigma = 0.1$. The acceptance and resolution functions are shown in Fig.1.
\begin{figure}[h]
\vspace*{-2cm}
\centerline{\hspace*{0.cm}\psfig{file=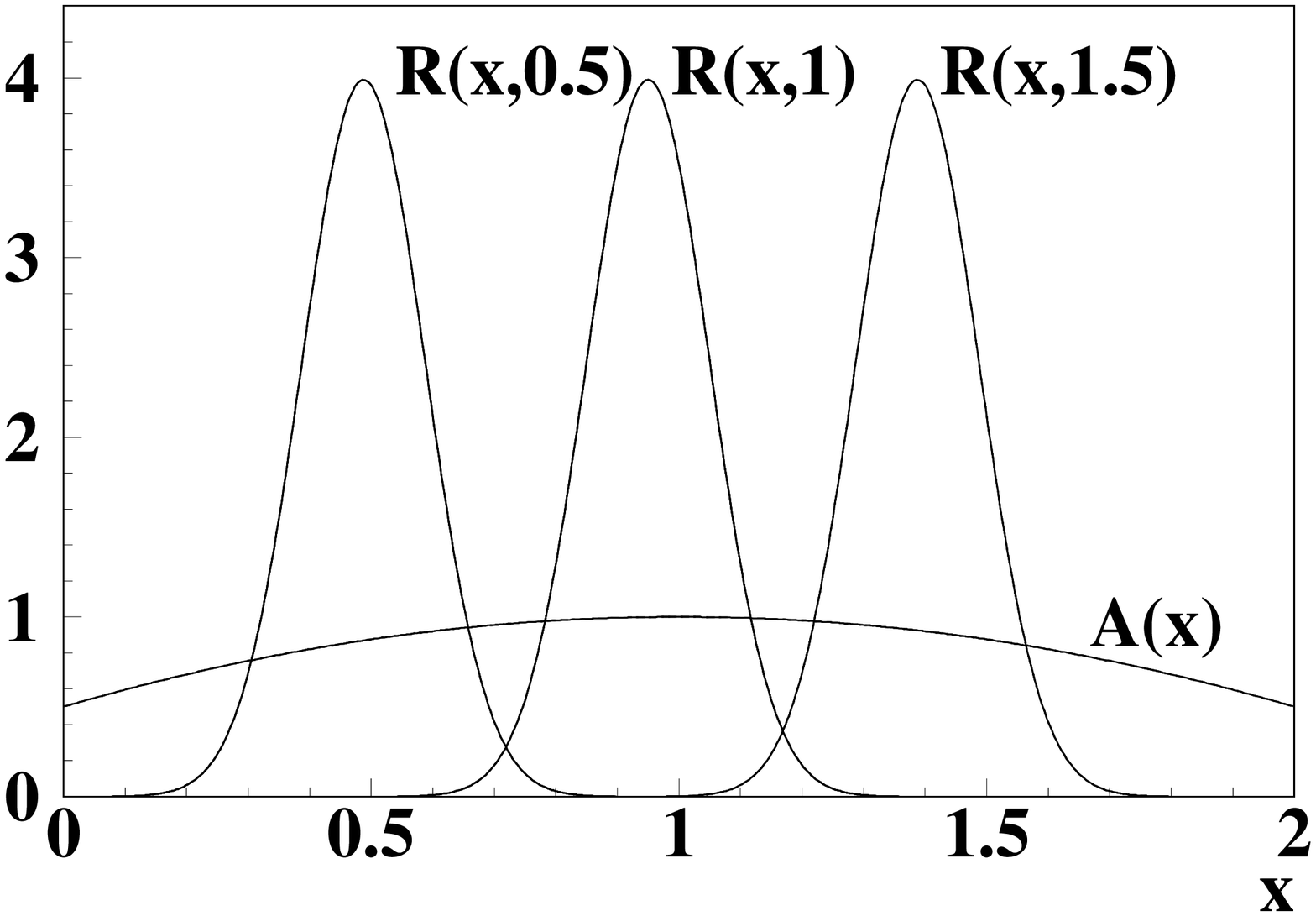,width=4.85in}}
\caption{The acceptance function $A(x)$ and resolution function $R(x,x')$ for $x'=0.5,  1.0$ and $1.5$.}
\end{figure}

\begin{figure}[h!]
\vspace*{-1. cm}
\centerline{\hspace*{0.5cm}\psfig{file=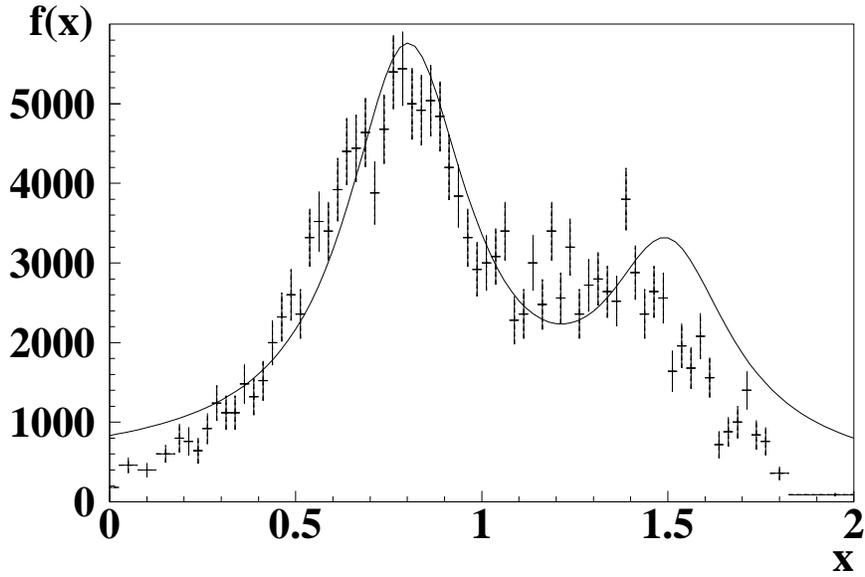,width=4.85in}}
\vspace*{-0.3cm}
\caption{The measured distribution $f(x)$ (number of events divided on bin size). The  true distribution $\phi(x)$ is shown as curve. }
\end{figure}

A histogram of the measured distribution $\bm{f}$ was obtained by simulating $5000$ events with $m=70$ bins, as shown in Fig. 2.

For the true distribution histogram, we chose $12$ bins of the same size as in a previous study \cite{blobel}. Fig. 3 shows  the histogram of the simulated true distribution. For detector identification we used a training sample comprising 100 distributions defined by formula (\ref{testform}) with parameters simulated according to uniform distributions on the intervals represented in Table 1.
Fig. 4 shows 50 of the 100 true distributions used for identification. Histograms of the measured distribution were obtained by generating $5000$ events.

\begin{figure}[h!]
\vspace*{-1cm}
\centerline{\hspace*{0.5cm}\psfig{file=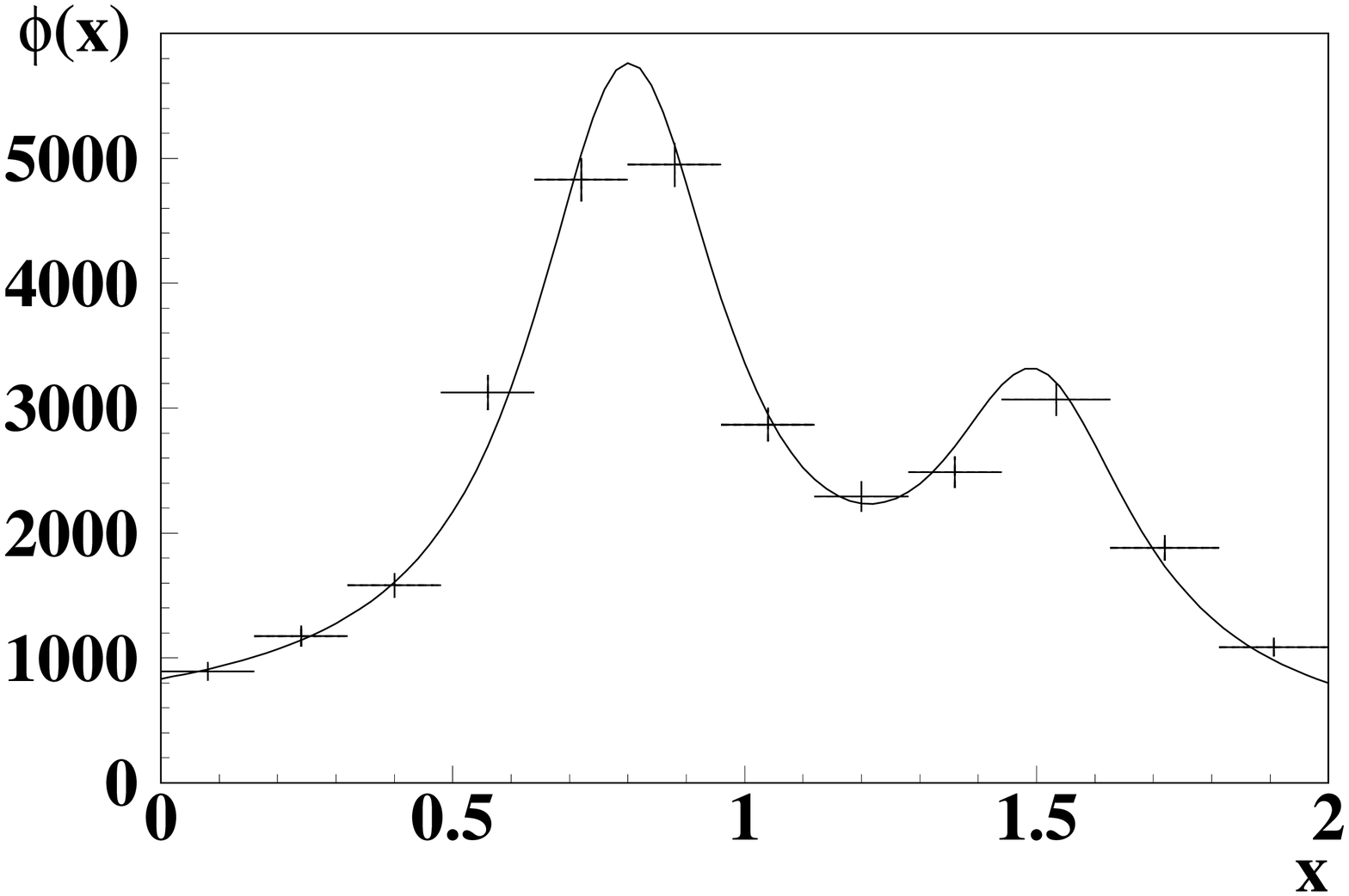,width=4.85in}}
\vspace*{-0.3cm}
\caption{The histogram of the simulated true distribution. The  true distribution $\phi(x)$ is shown as curve.}
\end{figure}
\begin{figure}[h!]
\vspace*{-1.cm}
\centerline{\hspace*{-0.0cm}\psfig{file=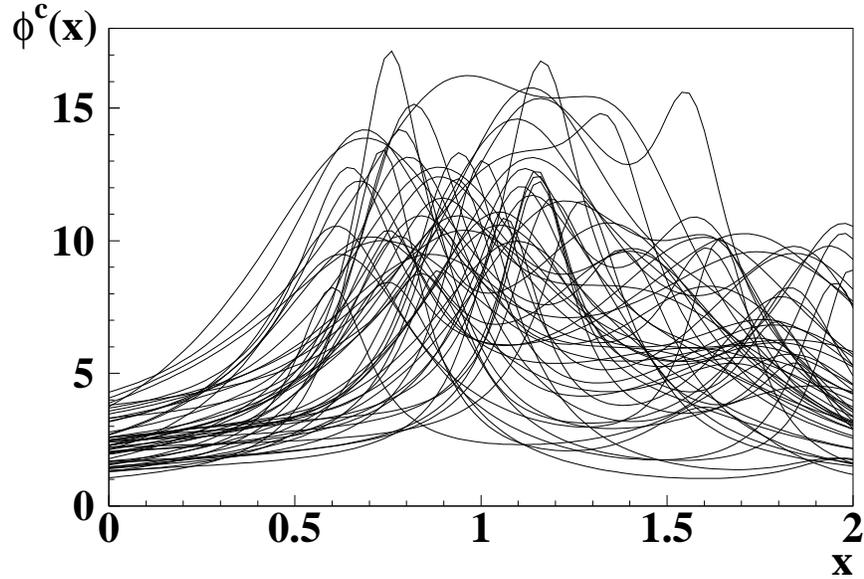,width=4.85in}}
\vspace*{-0.3cm}
\caption{The first 50 distributions for the training sample.}
\end{figure}

Matrix calculation was performed without $\mathrm{D}$-optimization. Table 2 shows the position of non-zero elements of matrix $P$. Elements of the matrix that are not close to elements defined by the greatest correlation are rather small. The maximum number of elements in each row of matrix $P$ that essentially defines the transformation is three. The matrix has approximately 20\% non-zero elements.

\begin{table}[h!]
\centering
\caption{Matrix $P'$, where ($\bullet$) denotes non-zero elements}
\tabcolsep 0.5pt
\scriptsize
\vspace *{0.3cm}
\begin{tabular}{cccccccccc cccccccccc cccccccccc cccccccccc cccccccccc cccccccccc cccccccccc}

$\bullet$ & $\bullet$ & $\bullet$ & $\bullet$ & $\bullet$ & o & o & o & o & o & o & o & o & o & o & o & o & o & o & o & $\bullet$ & o & o & o & o & o & o & o & o & o & $\bullet$ & o & o & o & o & o & o & o & o & o & o & o & o & o & o & o & $\bullet$ & o & o & o & o & o & o & o & o & o & o & o & o & o & o & o & o & o & o & o & o & o & o & o \\

$\bullet$ & o & o & o & o & $\bullet$ & $\bullet$ & $\bullet$ & o & o & o & o & o & $\bullet$ & o & o & o & o & o & o & o & o & o & o & o & o & o & o & o & o & o & o & o & o & o & o & o & o & o & o & o & o & o & o & o & o & o & o & o & o & o & o & o & o & o & o & o & o & o & $\bullet$ & o & o & o & o & o & o & o & o & o & o \\

 o & o & o & $\bullet$ & $\bullet$ & o & o & o & $\bullet$ & $\bullet$ & $\bullet$ & $\bullet$ & $\bullet$ & $\bullet$ & $\bullet$ & $\bullet$ & o & o & $\bullet$ & o & o & o & o & $\bullet$ & $\bullet$ & o & o & o & o & o & o & o & $\bullet$ & o & o & o & o & o & o & o & o & o & o & o & o & o & o & o & o & o & o & $\bullet$ & o & o & o & o & o & $\bullet$ & o & o & o & o & o & o & o & o & o & o & o & o \\

 o & o & o & o & o & o & o & o & o & o & o & o & $\bullet$ & $\bullet$ & $\bullet$ & $\bullet$ & $\bullet$ & $\bullet$ & $\bullet$ & $\bullet$ & $\bullet$ & $\bullet$ & $\bullet$ & o & o & o & o & o & $\bullet$ & o & o & o & o & o & o & o & o & o & o & o & o & o & o & o & o & o & o & $\bullet$ & o & o & o & o & o & o & o & o & o & o & o & o & o & o & o & o & o & o & o & o & o & o \\

 o & o & o & o & o & o & o & $\bullet$ & o & $\bullet$ & o & o & o & o & o & $\bullet$ & o & o & $\bullet$ & $\bullet$ & $\bullet$ & $\bullet$ & $\bullet$ & $\bullet$ & $\bullet$ & $\bullet$ & $\bullet$ & $\bullet$ & $\bullet$ & $\bullet$ & $\bullet$ & $\bullet$ & $\bullet$ & o & $\bullet$ & o & o & o & o & o & o & o & o & o & o & o & o & o & o & o & o & o & $\bullet$ & o & o & o & o & o & o & $\bullet$ & o & o & o & o & o & o & o & $\bullet$ & o & o \\

 o & o & o & $\bullet$ & $\bullet$ & $\bullet$ & o & o & o & o & o & o & o & o & o & o & o & o & o & o & o & o & o & o & $\bullet$ & $\bullet$ & $\bullet$ & $\bullet$ & $\bullet$ & $\bullet$ & $\bullet$ & $\bullet$ & $\bullet$ & $\bullet$ & $\bullet$ & $\bullet$ & $\bullet$ & $\bullet$ & o & o & o & o & o & o & o & o & o & o & $\bullet$ & o & o & o & o & o & o & o & o & o & o & o & o & o & o & o & o & o & o & o & o & o \\

 o & o & o & o & o & o & o & $\bullet$ & o & o & o & o & o & o & o & o & o & o & o & o & o & o & o & o & o & o & o & o & o & o & $\bullet$ & $\bullet$ & $\bullet$ & $\bullet$ & $\bullet$ & $\bullet$ & $\bullet$ & $\bullet$ & $\bullet$ & $\bullet$ & $\bullet$ & $\bullet$ & $\bullet$ & o & $\bullet$ & o & o & o & o & o & o & o & o & o & o & o & o & o & o & o & o & o & o & o & o & o & $\bullet$ & o & o & o \\
 o & o & o & o & o & o & o & o & o & o & o & o & o & o & o & o & o & o & o & o & o & o & o & o & o & o & o & o & o & $\bullet$ & o & o & o & o & $\bullet$ & $\bullet$ & $\bullet$ & $\bullet$ & $\bullet$ & $\bullet$ & $\bullet$ &
$\bullet$ & $\bullet$ & $\bullet$ & $\bullet$ & $\bullet$ & $\bullet$ & $\bullet$ & o & o & o & o & o & $\bullet$ & $\bullet$ & o & o & o & o & o & o & o
& o & o & o & $\bullet$ & o & o & o & o \\
o & o & o & o & o & o & o & o & o & o & o & o & o & o & o & o & o & o & o & o & o & o & o & o & o & o & o & o & o & o & o & o & o & o & $\bullet$ & o & o & o & o & o & o & $\bullet$ & $\bullet$ & o & $\bullet$ & $\bullet$ & $\bullet$ & $\bullet$ & $\bullet$ & $\bullet$ & $\bullet$ & $\bullet$ & $\bullet$ & o & o & o & o & o & o & o & o & o & o & o & o & o & o & o & o & o \\
$\bullet$ & o & o & o & o & o & o & o & o & o & o & o & o & o & o & o & o & o & o & o & o & o & o & o & o & o & $\bullet$ & o & o & o & o & o & o & o & $\bullet$ & o & o & o & $\bullet$ & $\bullet$ & o & o & o & $\bullet$ & o & $\bullet$ & $\bullet$ & $\bullet$ & o & $\bullet$ & $\bullet$ & $\bullet$ & $\bullet$ & $\bullet$ & $\bullet$ & $\bullet$ & $\bullet$ & $\bullet$ & $\bullet$ & $\bullet$ & $\bullet$ & o & o & o & o & o & o & o & o & o \\
o & o & o & o & o & o & o & o & o & o & o & o & o & o & o & o & o & o & o & o & o & o & o & o & o & o & o & o & o & o & o & o & o & o & o & o & o & o & o & o & o & o & o & o & o & o & o & o & o & o & o & o & $\bullet$ & o & $\bullet$ & $\bullet$ & $\bullet$ & $\bullet$ & $\bullet$ & $\bullet$ & $\bullet$ & $\bullet$ & $\bullet$ & $\bullet$ & $\bullet$ & $\bullet$ & $\bullet$ & o & o & o \\

o & o & o & $\bullet$ & o & o & o & o & o & o & o & o & $\bullet$ & o & o & $\bullet$ & o & o & o & o & o & o & o & o & o & o & $\bullet$ & o & o & $\bullet$ & o & o & $\bullet$ & o & o & o & o & o & o & o & o & o & o & o & o & o & o & o & o & o & o & o & o & o & o & $\bullet$ & o & o & $\bullet$ & $\bullet$ & $\bullet$ & $\bullet$ & $\bullet$ & $\bullet$ & $\bullet$ & $\bullet$ & $\bullet$ & $\bullet$ & $\bullet$ & $\bullet$\\

\end{tabular}
\end{table}

\begin{figure}
\vspace*{-2cm}
\centerline{\hspace*{0.5cm}\psfig{file=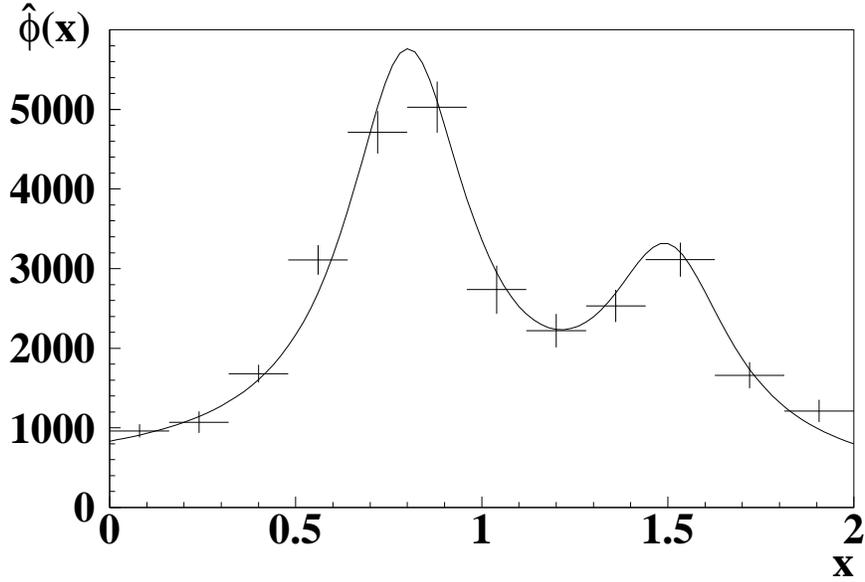,width=4.85in}}
\vspace*{-0.3cm}
\caption{The unfolded distribution $\hat{\phi}(x)$. The  true distribution $\phi(x)$ is shown as curve}
\end{figure}

Fig. 5 shows the unfolded distribution and the true distribution as a solid line. Comparison shows that the unfolded distribution basically reflects the fluctuations of the true distribution (see Fig. 3), but the statistical errors are greater. Table 3 presents errors and the correlation matrix for the unfolded distribution components. Errors are denoted as $\hat{\delta}_{ii}$ because they are only estimates of the error $\delta_{ii}$.

\begin{table}[h!]
 \centering
 \caption{Errors $\hat{\delta}_{ii}$ and correlation matrix for the unfolded distribution $\hat{\phi}(x)$}
 \vspace *{0.5 cm}
\footnotesize
\begin{tabular}{r|rrrrrrrrrrrrr}
 & $ \hat{\delta}_{ii} $ & 1 & 2 & 3 & 4 & 5 & 6 & 7 & 8 & 9 & 10 & 11 & 12 \\
 \hline
1 & 83 & & 0.3 & 0.0 & 0.0 & 0.0 & 0.0 & -0.1 & 0.0 & 0.0 & 0.1 & 0.0 & 0.0 \\
2 & 140 & 0.3 & & 0.1 & 0.0 & 0.0 & 0.1 & -0.1 & 0.0 & 0.0 & 0.0 & 0.0 & 0.0 \\
3 & 110 & 0.0 & 0.1 & & -0.1 & 0.0 & 0.0 & 0.0 & 0.0 & 0.0 & 0.0 & 0.0 & 0.0 \\
4 & 190 & 0.0 & 0.0 & -0.1 & & -0.3 & 0.1 & -0.1 & 0.0 & 0.0 & 0.0 & 0.0 & 0.0 \\
5 & 270 & 0.0 & 0.0 & 0.0 & -0.3 & & -0.5 & 0.3 & -0.1 & 0.1 & 0.0 & 0.0 & 0.0 \\
6 & 320 & 0.0 & 0.1 & 0.0 & 0.1 & -0.5 & & -0.6 & 0.3 & -0.1 & 0.0 & 0.0 & 0.0 \\
7 & 300 & -0.1 & -0.1 & 0.0 & -0.1 & 0.3 & -0.6 & & -0.5 & 0.0 & 0.1 & 0.0 & 0.0 \\
8 & 210 & 0.0 & 0.0 & 0.0 & 0.0 & -0.1 & 0.3 & -0.5 & & -0.2 & -0.1 & 0.1 & 0.0 \\
9 & 200 & 0.0 & 0.0 & 0.0 & 0.0 & 0.1 & -0.1 & 0.0 & -0.2 & & -0.4 & 0.1 & 0.0 \\
10 & 210 & 0.1 & 0.0 & 0.0 & 0.0 & 0.0 & 0.0 & 0.1 & -0.1 & -0.4 & & -0.4 & 0.2 \\
11 & 160 & 0.0 & 0.0 & 0.0 & 0.0 & 0.0 & 0.0 & 0.0 & 0.1 & 0.1 & -0.4 & & -0.3 \\
12 & 140 & 0.0 & 0.0 & 0.0 & 0.0 & 0.0 & 0.0 & 0.0 & 0.0 & 0.0 & 0.2 & -0.3 & \\
\end{tabular}
\end{table}

To investigate the statistical properties of the unfolding procedure, 1000 simulation runs were performed to produce 1000 statistically independent measured histograms for the same true distribution \ref{testform}. The unfolded distribution was calculated for each measured distribution. The same matrix $P$ was used for all cases. The following quantities were calculated:
\begin{itemize}

 \item Exact value of the components of the true distribution\\
 $\phi_i=5000\int_{x_i}^{x_{i+1}}\phi(x)dx/(x_{i+1}-{x_i})$ where $x_{i+1}$ and $x_i$ are the bounds of $i$th bin.
\item Average value of the components of the unfolded distribution\\ $\bar{\hat{\phi}}_i= \sum_{j=1}^{1000} \hat{\phi}_i(j)/1000$, where $j$ is the run number.
\item Bias for components of the unfolded distribution\\
 $B \hat{\phi}_i= \bar{\hat{\phi}}_i - \phi_i $
 \item Standard deviation $s_i$ for the unfolded distribution components\\
 $s_i= \sqrt {\sum_{j=1}^{1000}(\hat{\phi}_i(j)-\bar{\hat{\phi}}_i^2)/999}$.
 \item Mean estimated error $\hat{\delta}_{ii}$ for the unfolded distribution components \\ $\bar{\hat{\delta}}_{ii}= \sum_{j=1}^{1000} \hat{\delta}_{ii}(j)/1000$.
 \item Bias for errors in the unfolded distribution components\\
 $ B\hat{\delta}_{ii} = s_i-\bar{\hat{\delta}}_{ii}$.
\end{itemize}
\begin{table}[h!]
\caption{Exact values of components of the unfolded distribution $\phi_i$, average values $\bar{\hat{\phi}}_i$, bias $B \hat{\phi}_{i}$, standard deviation $s_i$, mean error $\bar{\hat{\delta}}_{ii}$ and bias for calculated errors $B\hat{\delta}_{ii}$}
\vspace *{0.5cm}
 \centering
 \begin{tabular}{r|rrrrrr}
$i$ & $\phi_i$ & $\bar{\hat{\phi}}_i$ & $B \hat{\phi}_i $& $s_i $ & $\bar{\hat{\delta}}_{ii}$ & $B\hat{\delta}_{ii}$\\
 \hline
1 & 913 & 968 & 55 & 82 & 84 & -2 \\
2 & 1152 & 1213 & 62 & 125 & 133 & -8 \\
3 & 1631 & 1666 & 36 & 117 & 116 & 1 \\
4 & 2760 & 2766 & 7 & 169 & 167 & 2 \\
5 & 4941 & 4897 & -44 & 265 & 252 & 12 \\
6 & 5011 & 4957 & -54 & 309 & 303 & 7 \\
7 & 3018 & 3070 & 53 & 292 & 298 & -6 \\
8 & 2284 & 2379 & 95 & 173 & 177 & -4 \\
9 & 2718 & 2770 & 53 & 205 & 199 & 6 \\
10 & 3073 & 2989 & -83 & 210 & 213 & -3 \\
11 & 1778 & 1776 & -3 & 160 & 162 & -2 \\
12 & 997 & 1037 & 40 & 129 & 138 & -9 \\
 \end{tabular}
\end{table}

The results presented in Table 4 and Fig. 6 show that the bias and statistical errors are small. Visual comparison of the unfolded distribution demonstrates the superiority of the present result over previous results \cite{blobel,hocker}. Comparison of biases is not possible because this has not been reported in any literature on unfolding methods.

\begin{figure}[h!]
\vspace*{-1cm}
\centerline{\hspace*{0.5cm}\psfig{file=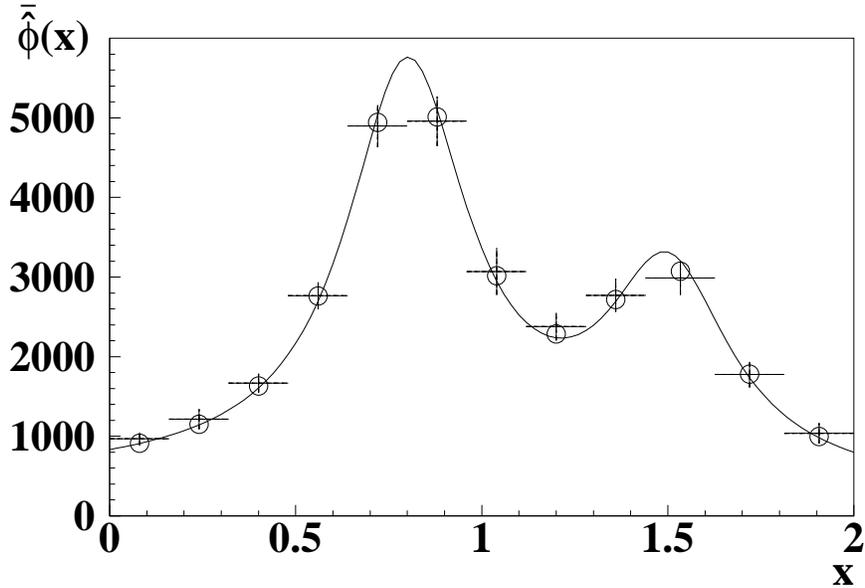,width=4.85in}}
\vspace*{-0.3cm}
\caption{Average values of components of the unfolded distribution $\bar{\hat{\phi}}_i$. Vertical bars denote the mean error $\bar{\hat{\delta}}_{ii}$ and circle centers ($\odot$) denote the exact value of components $\phi_i$. The  true distribution $\phi(x)$ is shown as curve}
\end{figure}
 \begin{table}[h!]
 \centering
 \caption{Values of the parameters chosen for numerical experiments}
 \vspace *{0.5cm}
 \hspace *{-0.7cm}
 \begin{tabular}{r|rrrrrrrrr}
 \textnumero & $A_1$ & $A_2$ & $A_3$ & $B_1$ & $B_2$ & $B_3$ & $C_1$ & $C_2$ & $C_3$ \\
 \hline
 1 & 1.38 & 8.85 & 5.19 & 0.52 & 0.93 & 1.79 & 1.61 & 0.30 & 0.36 \\ 2 & 0.56 & 13.18 & 4.05 & 0.22 & 0.78 & 1.91 & 1.26 & 0.16 & 0.45 \\
 3 & 0.55 & 9.97 & 5.67 & 0.32 & 1.20 & 1.64 & 3.02 & 0.35 & 0.20 \\ 4 & 2.21 & 9.28 & 3.79 & 0.48 & 0.67 & 1.61 & 2.11 & 0.20 & 0.48 \\
 5 & 2.77 & 9.02 & 7.61 & 0.49 & 0.64 & 1.72 & 3.35 & 0.17 & 0.41 \\
 6 & 1.66 & 7.94 & 1.18 & 0.39 & 1.09 & 1.43 & 2.44 & 0.30 & 0.30 \\
 7 & 1.19 & 9.06 & 6.88 & 0.51 & 1.06 & 1.81 & 1.87 & 0.11 & 0.25 \\
 8 & 1.31 & 7.13 & 7.97 & 0.31 & 0.77 & 1.41 & 3.23 & 0.17 & 0.34 \\
 \end{tabular}
\end{table}

To demonstrate that the algorithm is robust, eight sets of parameters (Table 5) were randomly simulated according to uniform distributions on the intervals represented in Table 1. For each set, a random experiment with 1000 runs was performed using matrix $P$ defined in the first example. The results presented in Fig. 7 demonstrate the robustness of the method, with rather low bias for the unfolded distribution in all eight cases.

\begin{figure}[h!]
\vspace*{-7.5cm}
\centerline{\hspace*{2.7cm}\psfig{file=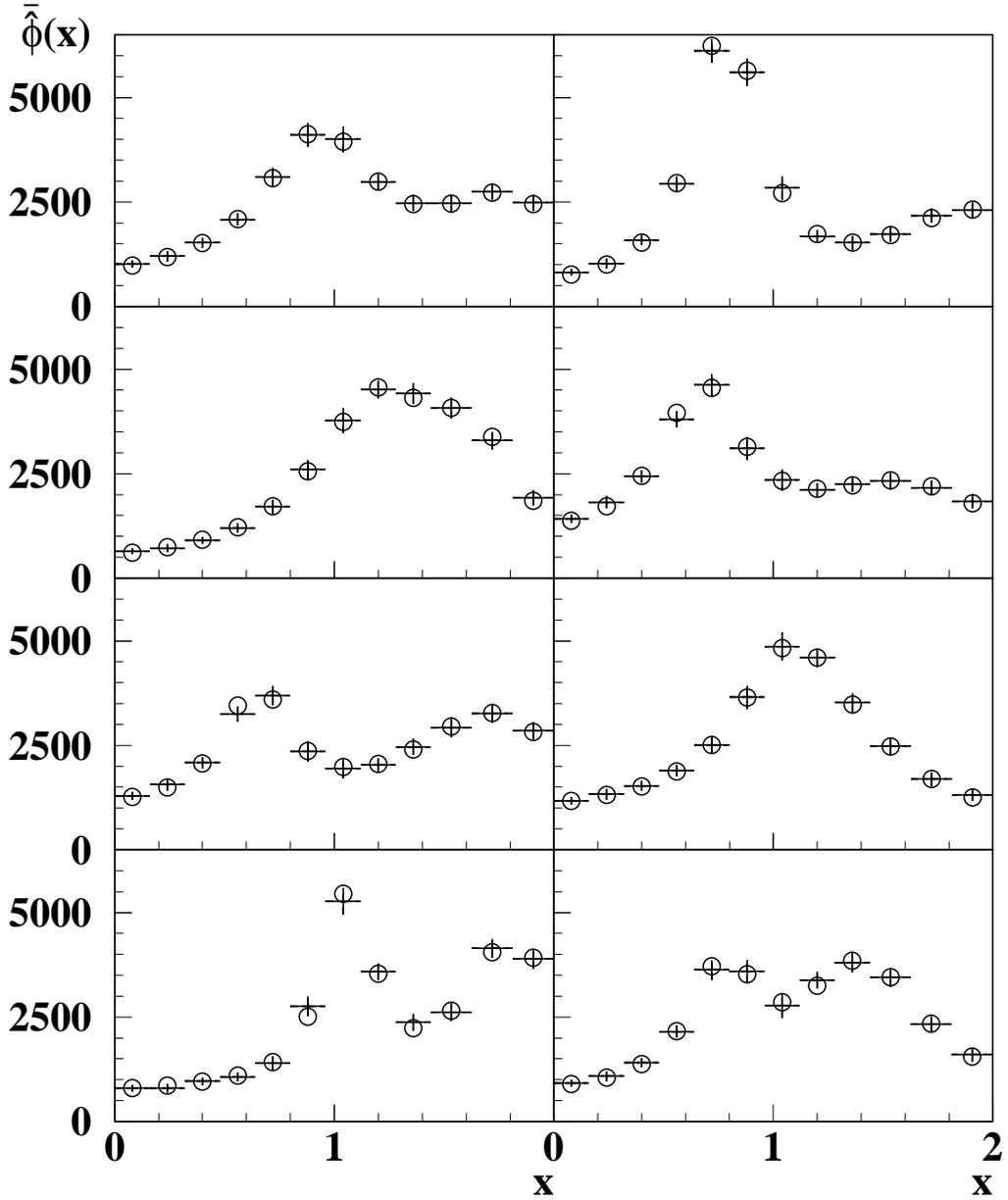
,width=6.5in}}
\vspace*{-0.4cm}
\caption{Results of numerical experiments for the eight distributions defined in Table 5 from left to right and top to bottom. Graphs show average values of components of the unfolded distributions $\bar{\hat{\phi}}_i$. Vertical bars denote the mean error $\bar{\hat{\delta}}_{ii}$, and circle centers ($\odot$) denote the exact value of components $\phi_i$.}
\end{figure}

\newpage

\section{Discussion and conclusion}

The main difficulties of the unfolding problem, which is a particular case of the inverse problem, are widely known. Information is lost in measuring owing to the inefficiency of registration in the frequency domain because of the low-pass filter defined by the resolution function and to the inefficiency of events registration defined by the acceptance function of the set-up. Thus, there are an infinite number of true distributions that give the same measured distribution and therefore a priori information about the solution must be used to solve an unfolding problem (inverse problem).

 One way to solve an unfolding problem is to replace the original problem by a problem for a smoothed original true distribution and to use a sliding window (bin) for a smoothing. This is equivalent to solving the unfolding problem for the true distribution in some binning. Smoothing is low-pass filtering and the loss of information for a smoothed distribution due to the resolution function effect, which is another low-pass filter, is lower than for the original true distribution. Solution of the unfolding problem is easier, but no information is obtained about the structure of the original true distribution inside the bin.

In practical applications of the unfolding procedure, the transformation matrix $P$ must be calculated. Simulation of the measurement process is used for this, especially in nuclear and particle physics. This process is very time-consuming and the sample size for simulated events is often of the same order as for measured events. The calculated matrix will have many noisy matrix elements in this case, which is another source of instability in solving the inverse problem.

Main points related with difficulties of the unfolding problem have formulated above on physical level of rigor permit us summarize results of given paper and define place of proposed unfolding method among other known methods.

The method presented here is a completely new approach to unfolding problems using machine learning concepts, including a training sample, a validation procedure and boosting. All a priori information about the solution is contained in the training sample, which is a set of physically motivated true distributions known from theory and other experiments. Methods for selecting distributions for the training sample were presented in Section 3 and are supported by previous research \cite{goodness,gagunashviliph}.

In the proposed method, an unfolded distribution can be calculated for a grid of points or for bins. There are no restrictions imposed by the dimensionality of the problem or the configuration of the bins or the grid. The method for identification provides a linear approximation of a transformation from the true distribution to the measured distribution if this transformation is non-linear.

The numerical example presented demonstrates the robustness of the new unfolding procedure and the possibility of unfolding a whole set of distributions with a single calculated matrix for transformation $P$. The set is defined as distributions used to create the training sample. Biases and statistical errors for components of the unfolded distribution were calculated using a Monte-Carlo method with 1000 runs. The examples demonstrate that the bias is small for components of the unfolded distribution and for estimates of the statistical errors. It should be noted that such biases were investigated for unfolding for the first time.
The unfolding procedure is validated using a machine learning approach and has a good statistical interpretation.
The proposed method has wide potential for applications in nuclear and particle physics, where  models for training samples can be proposed and Monte-Carlo simulations can be used to calculate transformation matrices.

\end{document}